\documentclass[graybox]{svmult}

\usepackage{helvet}         
\usepackage{courier}        
\usepackage{type1cm}        
\usepackage{makeidx}         
\usepackage{graphicx}        
\usepackage{multicol}        
\usepackage[bottom]{footmisc}

\usepackage{subcaption}
\usepackage{amsmath}

\makeindex             

\begin{document}

\title*{Simulated Charge Stability in a MOSFET Linear Quantum Dot Array}
\titlerunning{Charge Stability in a Linear Quantum Dot Array}

\author{Zach D. Merino, Bohdan Khromets, and Jonathan Baugh}
\authorrunning{Zach D. Merino and Jonathan Baugh}
\institute{Zach D. Merino
    \at University of Waterloo, ON, N2L 3G1, 
    \email{zmerino@uwaterloo.ca}
\and Bohdan Khromets 
    \at University of Waterloo, ON, N2L 3G1, 
    \email{bohdan.khromets@uwaterloo.ca}
\and Jonathan Baugh 
    \at Institute for Quantum Computing
    \at University of Waterloo, ON, N2L 3G1 \email{baugh@uwaterloo.ca}}
%
%
\maketitle

\abstract{
In this study, we address challenges in designing quantum information processors based on electron spin qubits in electrostatically-defined quantum dots\index{Quantum Dots} (QDs). Numerical calculations of charge stability diagrams are presented for a realistic double QD device geometry. These methods generalize to linear QD arrays, and are based on determining the effective parameters of a Hubbard model\index{Hubbard Model} Hamiltonian that is then diagonalized to find the many-electron ground state energy. These calculations enable the identification of gate voltage ranges that maintain desired charge states during qubit manipulation, and also account for electrical cross-talk between QDs. As a result, the methods presented here promise to be a valuable tool for developing scalable spin qubit quantum processors.}

\section{Introduction}
\label{sec:1}

In recent years, certain Quantum Computing\index{Quantum Computing} (QC) implementations  have demonstrated a quantum advantage over classical processors \cite{Zhong2020}. While further demonstrations of quantum supremacy may be realized, quantum processors (QPs) consisting of thousands to millions of 
physical qubits will be required to perform large-scale fault-tolerant QC.

\begin{figure}[htb!]
    \centering
    \begin{subfigure}[h!]{0.3\textwidth}
        \centering
        \includegraphics[width=\linewidth]{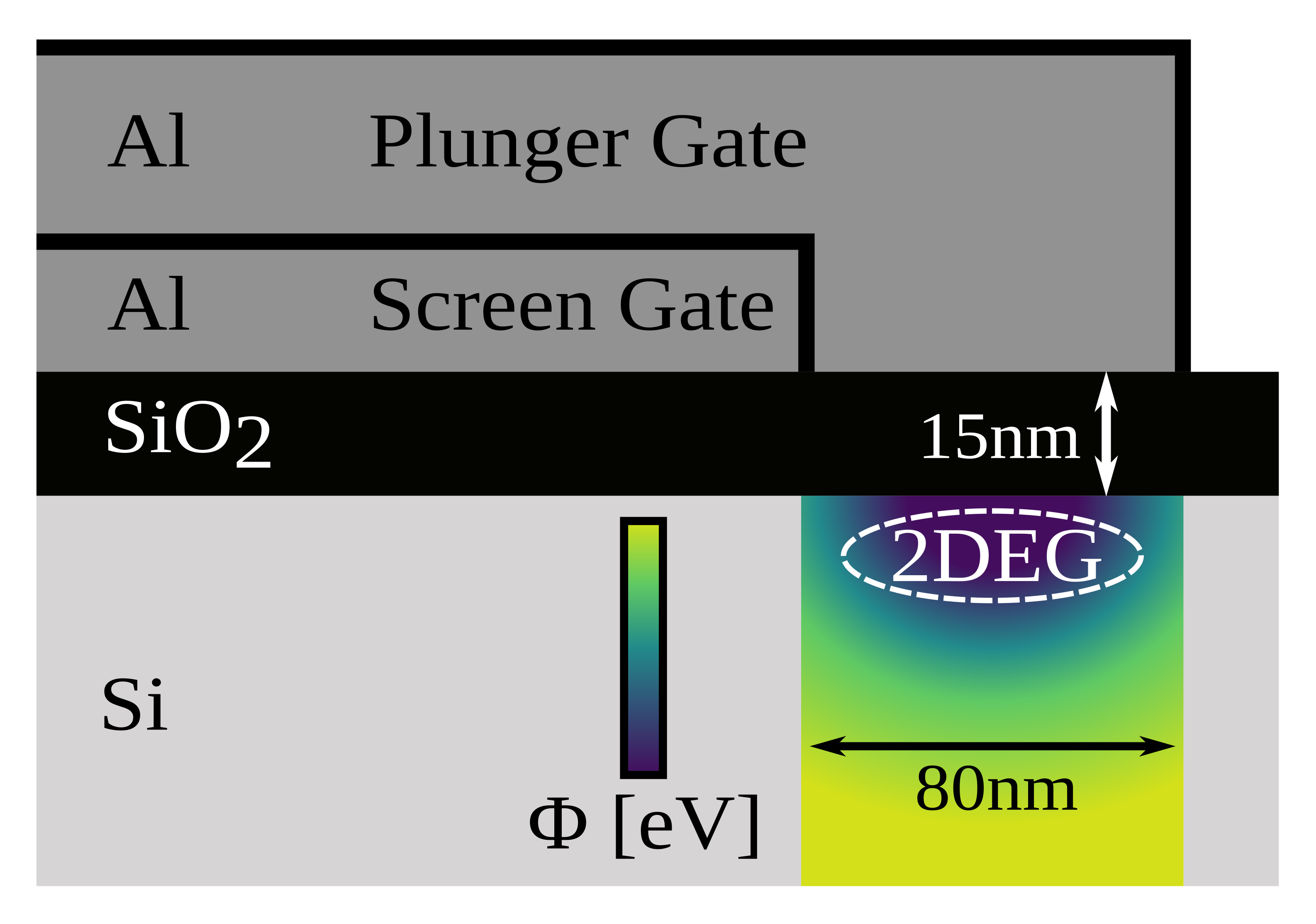} 
        \caption{} \label{fig: fig1a}
    \end{subfigure}
    \hspace{0.1em}%
    \begin{subfigure}[h!]{0.35\textwidth}
        \centering
        \includegraphics[width=\linewidth]{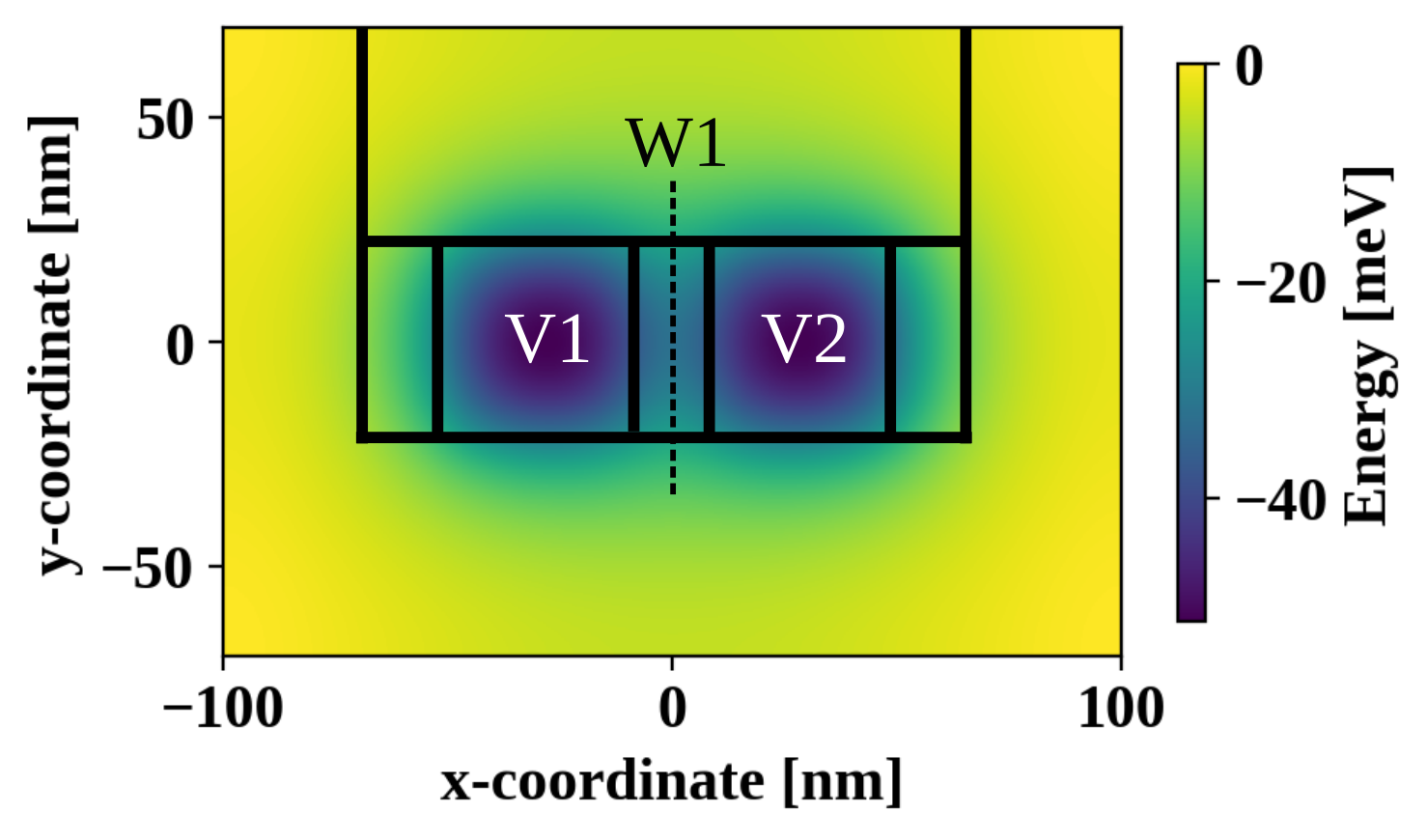} 
        \caption{} \label{fig: fig1b}
    \end{subfigure}
    \hspace{0.1em}%
    \begin{subfigure}[h!]{0.3\textwidth}
        \centering
        \includegraphics[width=\linewidth]{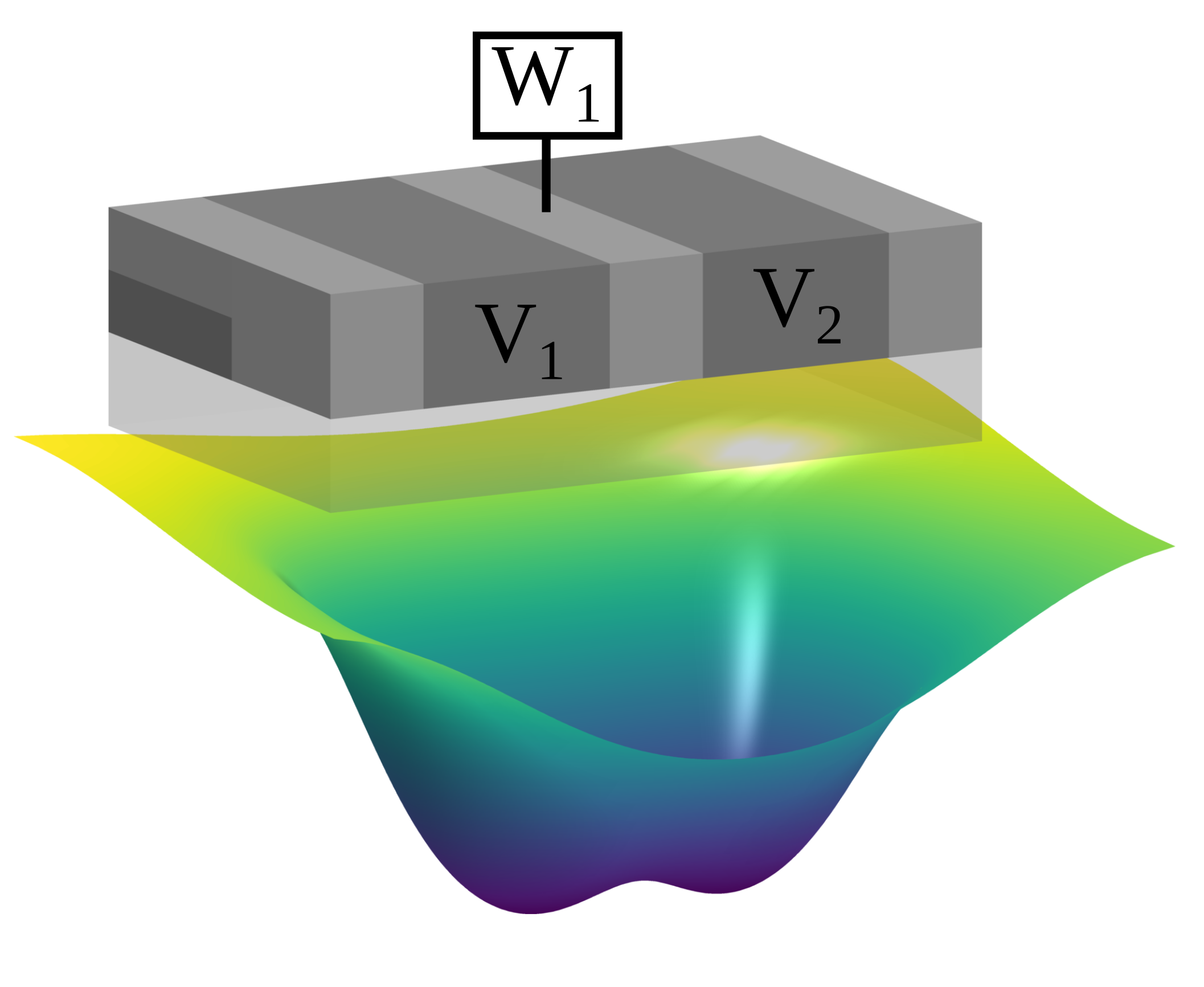} 
        \caption{} \label{fig: fig1c}
    \end{subfigure}
    \caption{A schematic MOSFET double QD device in Si: (\subref{fig: fig1a}) Side cross-sectional view showing a 2-dimensional electron gas (2DEG) formed by a plunger gate electrode.
    (\subref{fig: fig1b}) Top-view cross-section showing the double well confining potential formed by the plunger ($V_1, V_2$) and tunnel ($W_1$) electrodes. The electrostatic potential energy at the SiO\textsubscript{2}/Si interface is calculated by a finite element Poisson solver based on a 3D device model. (\subref{fig: fig1c}) Contour plot of the potential energy landscape in (b).} \label{fig:fig1}
\end{figure}

The scaling requirements for fault tolerant QC make electrostatically-defined quantum dot (QD) 
spin qubits a promising approach. In particular, spin qubits in silicon (Si) have attracted much recent interest due to their compatibility with industrial complementary metal-oxide semiconductor (CMOS) fabrication techniques, which could yield high quality, densely packed qubits \cite{Saraiva2021, Undseth2023}. A schematic Si double QD device is shown in figure \ref{fig:fig1}. A 2-dimensional electron gas accumulates at the Si/SiO$_2$ interface due to positive voltage applied to a gate electrode. Mesoscopic QDs are formed beneath the plunger gate fingers, localized by the presence of screening gates. The QDs are tuned into the few-electron regime via the plunger gates, where tunnel coupling to a reservoir (not shown) allows for charge transfer. The spin of an unpaired electron in a QD with an odd number of charges constitutes the qubit. Spin qubits in Si and SiGe devices have exhibited long coherence compared to quantum gate operation time scales \cite{Veldhorst2014}, high-fidelity gate operations required for fault-tolerant QC \cite{Mills2022}, qubit control via applied electric gate signals \cite{Undseth2023}, 
and qubit operation at relatively high temperatures ($1-4$~K) \cite{Petit2020}.
Furthermore, scalable spin qubit architectures suitable for implementing surface code quantum error correction have been proposed \cite{Veldhorst2017, Buonacorsi2019}.

Scaling QPs generally increases the complexity in initializing, controlling, and measuring quantum computational states. Each additional QD adds independent control parameters (plunger and tunnel gates) and is a source of cross-talk between neighboring QDs \cite{Heinz2021, Undseth2023}. A critical requirement is the ability to predict the charge state of a linear QD array for a given vector of gate voltages. For spin qubits in singly-occupied QDs, it is necessary to determine the control boundaries within which single charge occupations are maintained. Thus, accurately modeling the charge state of a realistic linear QD array as a function of gate voltages is a necessary first step, prior to the design of spin qubit control pulses. 

Previous studies have employed Hubbard models as approximate descriptions of linear arrays of tunnel-coupled QDs  \cite{Hensgens2017, Secchi2023}. Given certain assumptions about the underlying electrostatic potential, others have shown that effective parameters of the Hamiltonian can be 
calculated \cite{DasSarma2011, Yang2011} and a resulting 
charge configuration determined \cite{Jafari2008}. In this work,  electrostatic potential landscapes for realistic QD device geometries found by a finite element Poisson solver \cite{Birner2007} are used as inputs for calculations that determine stable charge states as a function of gate electrode voltages. 

\section{Hubbard Model for Quantum Dots}
\label{sec:2}

\subsection{Hamiltonian}
\label{sec:2.1}

A generalized Hubbard Hamiltonian developed by \cite{DasSarma2011, Yang2011} for electrostatically defined QD systems is composed of
\begin{equation}
	H=H_{\mu}+H_{t}+H_{U}+H_{J},\label{eq:hubbard_ham}
\end{equation} 
\noindent where terms describing the chemical potential, $H_{\mu}$, tunnel coupling, $H_{t}$, Coulomb repulsion, $H_{U}$, and spin exchange, $H_{J}$, are included. The first term, $H_{\mu}$, represents the occupation of spin states in an array of $N$ quantum dots:
\begin{equation} \label{eq:H_chem_pot}
	H_{\mu}=-\sum_{\sigma =\uparrow, \downarrow} 
    \left[
        \sum_{i=1}^{N}\mu_{i}
        \hat{n}_{i,\sigma}
    \right]
\end{equation}
\noindent Here, $i$ is the index of each QD in a linear array of $N$ QDs, $\mu_i$ are their chemical potentials, $\sigma \in \{ \uparrow, \downarrow\}$ is the spin state, and $\hat{n}_{i,\sigma}$ is the electron number operator for the spin state $\sigma$ of the $i$\textsuperscript{th} QD. The number operator is defined as 
$\hat{n}_{i,\sigma}=\hat{c}_{i,\sigma}^{\dagger}\hat{c}_{i,\sigma}$, where $\hat{c}_{i,\sigma}^{\dagger}$ $(\hat{c}_{i,\sigma})$ are the usual creation (annihilation) operators. 

The second term, $H_{t}$, represents the tunnel couplings between QDs where an electron with spin state $\sigma$ is allowed to tunnel to an adjacent QD.
\begin{equation} \label{eq:H_hopping}
	H_{t}=- \sum_{\sigma =\uparrow, \downarrow} 
    \left[
    \sum_{j=1}^{N-1}t_{j,j+1}
        \left(\hat{c}_{j,\sigma}^{\dagger}\hat{c}_{j+1,\sigma}
            +\hat{c}_{j+1,\sigma}^{\dagger}\hat{c}_{j,\sigma}\right)
    \right]
\end{equation}

\noindent Here, $t_{j,j+1}$ are the tunnel coupling, or inter-site hopping terms for adjacent QDs. The operators $\hat{c}_{j,\sigma}^{\dagger}\hat{c}_{j+1,\sigma}$, $\left(\hat{c}_{j+1,\sigma}^{\dagger}\hat{c}_{j,\sigma}\right)$, account for $j \rightarrow (j+1)$, $\left(j \leftarrow (j+1)\right)$, tunneling events.


The third term, $H_{U}$, is the Coulomb repulsion between all electron pairs.
\begin{equation} \label{eq: H_U}
	H_{U} =  \sum_{i = 1}^{N} U_{i} 
                    \hat{n}_{i,\uparrow}\hat{n}_{i,\downarrow} \nonumber
        + \sum_{\sigma_1,\sigma_2 \in \{\uparrow, \downarrow\}}
        \left[\sum_{j=1}^{N-1} 
            U_{j,j+1}
            \hat{n}_{j,\sigma_1}\hat{n}_{j+1,\sigma_2}
        \right]
\end{equation}

The intra-Coulomb, or onsite, energy, $U_i$, accounts for the Coulomb repulsion energy of electrons on the same QD, while the inter-Coulomb energy, $U_{j,j+1}$, accounts for the Coulomb energy between pairs of electrons in adjacent QDs. In general, the inter-Coulomb energy accounts for all combinations of electron pairs in different QDs, however, in this work only nearest neighbor contributions are considered.

The last term, $H_{J}$, accounts for energy contributions from spin-exchange, pair-hopping, and occupation-modulated hopping. However, this term is much smaller compared to the other three components and will be neglected for the purposes of this work.

\subsection{Effective Parameters}
\label{sec:2.2}

\begin{figure}[ht]
    \centering
    \begin{subfigure}[h!]{0.45\textwidth}
        \centering
        \includegraphics[width=\linewidth]{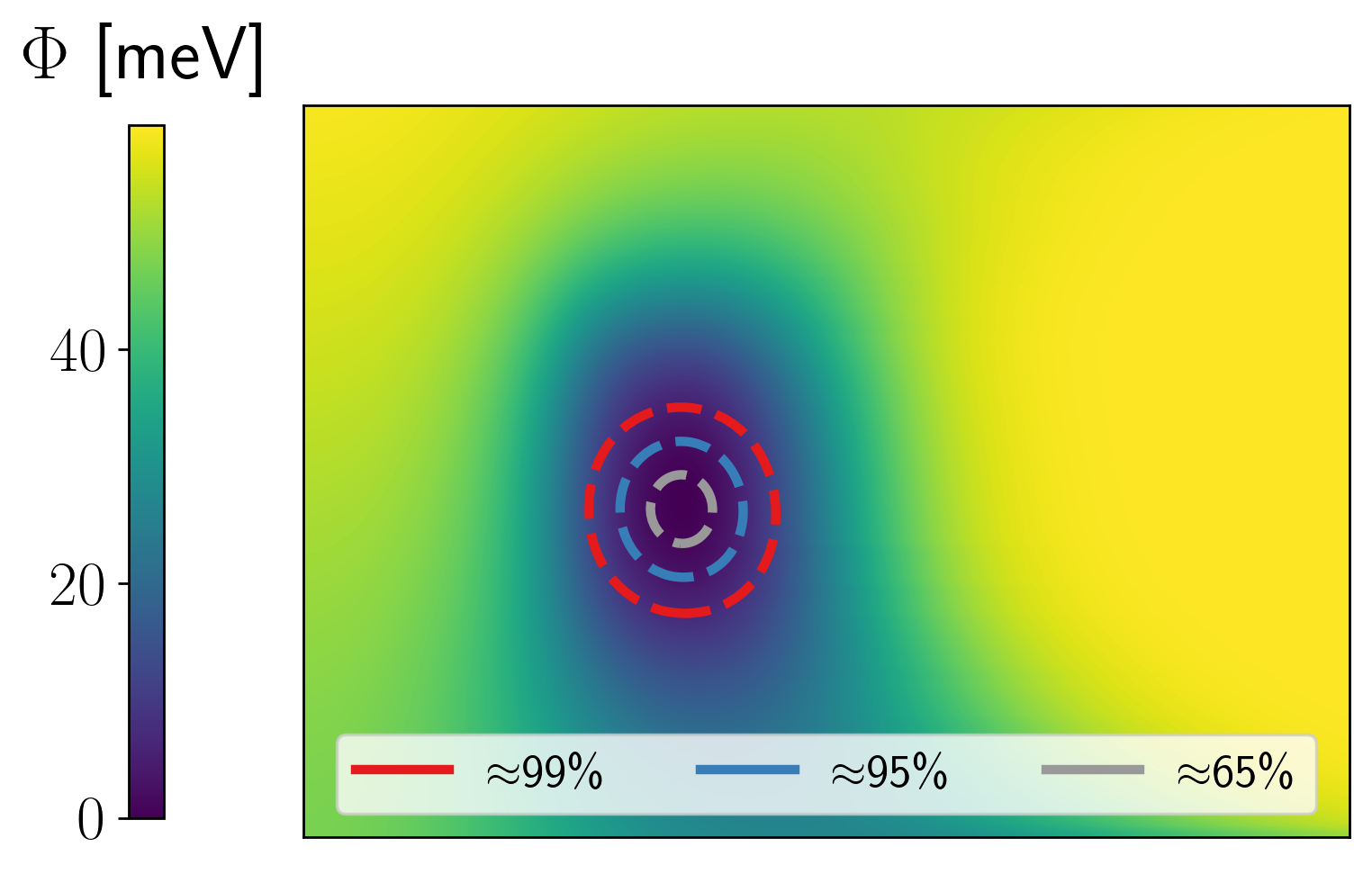} 
        \caption{} \label{fig: fig2a}
    \end{subfigure}
    \hspace{0.1em}%
    \begin{subfigure}[h!]{0.45\textwidth}
        \centering
        \includegraphics[width=\linewidth]{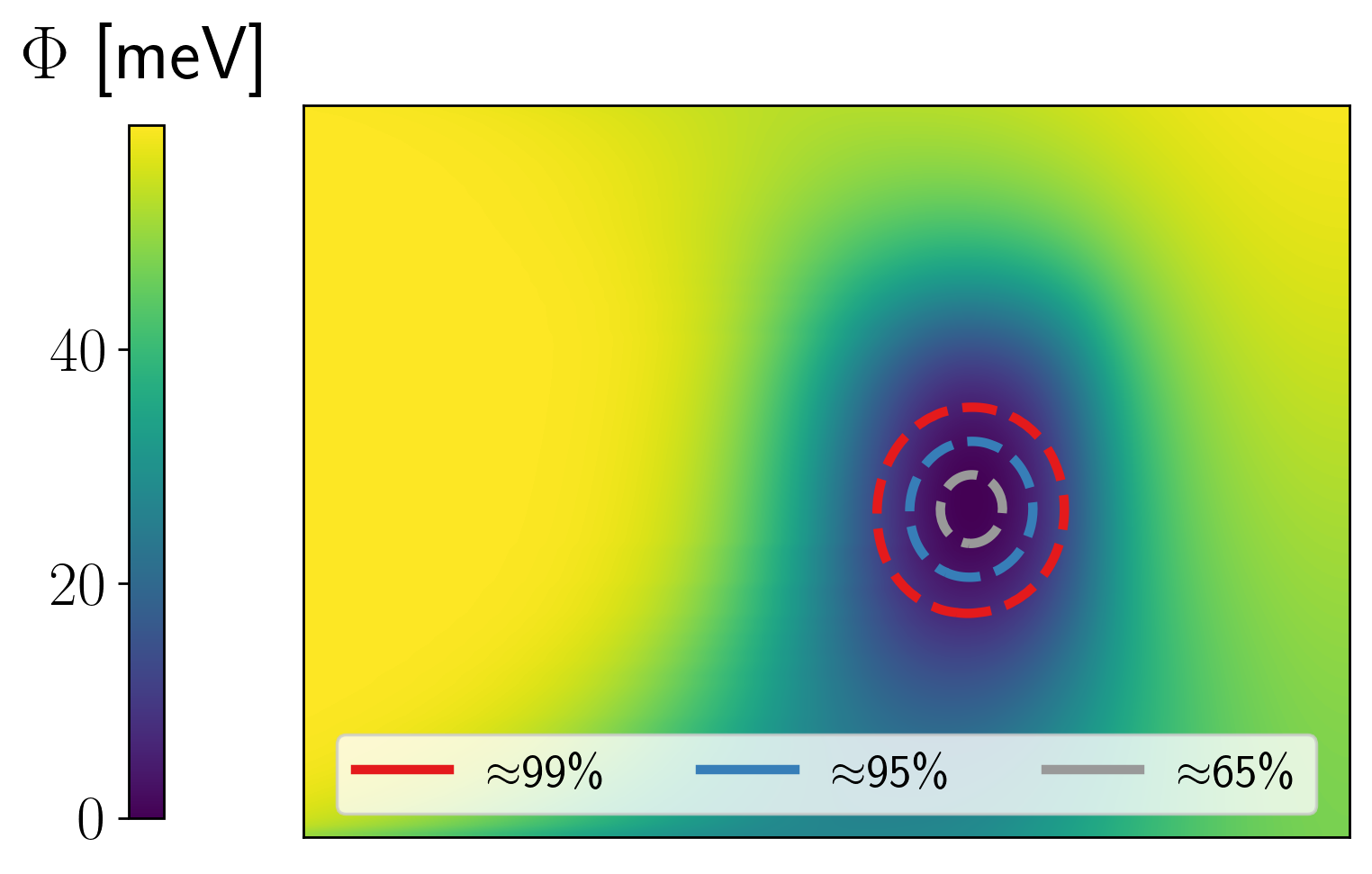} 
        \caption{} \label{fig: fig2b}
    \end{subfigure} 
    \caption{Electrostatic potential, $\Phi$, for:
    (\subref{fig: fig2a}) left single QD and masked right QD
    (\subref{fig: fig2b}) right single QD and masked left QD. Regions are in circled where there is a $\approx 99\%$, $\approx 95\%$, and $\approx 65\%$ probability of finding an electron.} \label{fig: fig2}
\end{figure}

The inter-Coulomb repulsion is calculated using the expression given in 
\cite{Yang2011},
\begin{equation} \label{eq:Uij}
	U_{ij}=\int_{-\infty}^{\infty} d\boldsymbol{r_1}
    \int_{-\infty}^{\infty} d\boldsymbol{r_2}
    \left|\varphi_{i}(\boldsymbol{r_1})\right|^{2}
    \left|\varphi_{j}(\boldsymbol{r_2})\right|^{2}
    v(\boldsymbol{r_1}, \boldsymbol{r_2})
\end{equation}

\noindent where $\varphi_{i}$ is the single-particle ground state wave function 
of the $i$\textsuperscript{th} QD and,
\begin{equation} \label{eq:v}
    v(\boldsymbol{r_1}, \boldsymbol{r_2}) \equiv
        \frac{e^2}{4 \pi \epsilon_0 \epsilon_r}\frac{1}{\left|\boldsymbol{r_1}-\boldsymbol{r_2}\right|} \nonumber
\end{equation}
\noindent is the Coulomb repulsion term. The onsite repulsion term for the $i$\textsuperscript{th} QD is defined likewise but contains only the single-particle wave function $\varphi_{i}$ of the QD in consideration:
\begin{equation} \label{eq:Ui}
	U_{i} = U_{ii}
\end{equation} The wave functions 
$\varphi_{i}$ are determined by solving the Schrödinger equation 
\begin{align} \label{eq:schrodinger}
    \left[ 
       \frac{\boldsymbol{\hat{p}}_i^2 }{2m^*} + V_{i} 
        \left( \boldsymbol{r} \right) 
    \right] \varphi_i \left( \boldsymbol{r} \right) 
    = \varepsilon_i \varphi_i \left( \boldsymbol{r} \right),
\end{align}
\noindent
where $m^*$ is the electron effective mass, 
for all electrostatic potentials 
$V_{i} \left( \boldsymbol{r} \right)$ representing localized individual QDs. An example of the procedure that allows us to determine $V_{i} \left( \boldsymbol{r} \right)$ for each QD in a double-dot potential is shown in figure \ref{fig: fig2}. Here, each of the regions near the local minima (describing individual QDs) is fitted to a 2D Gaussian function while the other ones are masked with a high potential value.
The procedure preserves the effects of gate voltage cross-talk on QDs even after the isolation of their electrostatic potentials.

\noindent While the equations (\ref{eq:Uij}) and (\ref{eq:Ui}) are well defined, the integral expressions appear to be singular due to the Coulomb potential term $v(\boldsymbol{r_1}, \boldsymbol{r_2})$ when $\boldsymbol{r_1} = \boldsymbol{r_2}$. However, the apparent singularity can be avoided by performing the 
change of variable
\begin{align*}
	\boldsymbol{r} = \boldsymbol{r_1} - \boldsymbol{r_2}
\end{align*}
and integrating $\boldsymbol{r}$ in polar coordinates. This yields the following integral that can be performed 
numerically:
\begin{align} \label{eq:Uij_polar}
	U_{ij}= \frac{e^2}{4 \pi \epsilon_0 \epsilon_r} \int_{0}^{\infty} d\left|\boldsymbol{r}\right|
    \int_{0}^{2 \pi} d\theta
    \int_{-\infty}^{\infty} d\boldsymbol{r_2}
    \left|\varphi_{i}(\boldsymbol{r} + \boldsymbol{r_2})\right|^{2}
    \left|\varphi_{j}(\boldsymbol{r_2})\right|^{2}
\end{align}

Next, to determine the chemical potential from equation (\ref{eq:H_chem_pot}), $V_{i} \left( \boldsymbol{r} \right)$ is defined by masking all but the $i$\textsuperscript{th} QD's electrostatic potential, and the single-electron ground state energy is found by solving equation (\ref{eq:schrodinger}).
\begin{align*}
    \mu_i = \varepsilon^{(g)}_i
\end{align*}

Lastly, the tunnel coupling term from equation (\ref{eq:H_hopping}) is derived from a two-level Hamiltonian that includes interdot tunneling in a double QD, and by masking the electrostatic potential for all but the $j$\textsuperscript{th} QD pair, 
\begin{equation}
    t_{j, j+1} = \frac{1}{2} 
        \sqrt{
            \Bigr( E^{(e)}_{j, j+1} - E^{(g)}_{j, j+1} \Bigl)^2
            + \Bigr( \varepsilon^{(g)}_{j} - \varepsilon^{(g)}_{j+1} \Bigl)^2
            }
\end{equation}

\noindent where $E^{e/g}_{j,j+1}$ are the first excited/ground state energies for the double QD potential of the $j$\textsuperscript{th} QD pair and $\varepsilon^{(g)}_{j/j+1}$ are the ground state energies corresponding to the single QD potentials. 

\subsection{Charge State via Exact Diagonalization}
\label{sec:2.3}

To determine the charge state of a linear QD array at a given voltage configuration, 
the components of the Hamiltonian defined by equations 
(\ref{eq:H_chem_pot})-(\ref{eq: H_U}) are calculated for a chosen charge 
configuration basis. The Hubbard model \eqref{eq:hubbard_ham} accounts for all charge and spin states with up to two electrons per QD following the Pauli exclusion principle. For example, a choice of basis for a double QD with total electron number 2 is the following:

\begin{equation*}
    \begin{aligned}[c]
        \left|\alpha_1\right\rangle  
        & \equiv c_{1,\uparrow}^{\dagger}c_{2,\uparrow}^{\dagger}\left|0,0\right\rangle 
        =\left|\uparrow,\uparrow\right\rangle \\
	\left|\alpha_2\right\rangle  
        & \equiv c_{2,\downarrow}^{\dagger}c_{2,\uparrow}^{\dagger}\left|0,0\right\rangle 
        =\left|0,\uparrow\downarrow\right\rangle \\
	\left|\alpha_3\right\rangle  
        & \equiv c_{1,\downarrow}^{\dagger}c_{2,\uparrow}^{\dagger}\left|0,0\right\rangle 
        =\left|\downarrow,\uparrow\right\rangle
    \end{aligned}
    \qquad
    \begin{aligned}[c]
        \left|\alpha_4\right\rangle  
        & \equiv c_{2,\downarrow}^{\dagger}c_{1,\uparrow}^{\dagger}\left|0,0\right\rangle 
        =\left|\uparrow,\downarrow\right\rangle \\
	\left|\alpha_5\right\rangle  
        & \equiv c_{1,\downarrow}^{\dagger}c_{1,\uparrow}^{\dagger}\left|0,0\right\rangle 
        =\left|\uparrow\downarrow, 0 \right\rangle \\
	\left|\alpha_6\right\rangle  
        & \equiv c_{1,\downarrow}^{\dagger}c_{2,\downarrow}^{\dagger}\left|0,0\right\rangle 
        =\left|\downarrow,\downarrow\right\rangle
    \end{aligned}
    \end{equation*}

\noindent The ordering of the basis is arbitrary, however, the above order was chosen for consistency with \cite{Jafari2008} so that the Hamiltonian has only diagonal contributions from $H_{\mu}$ and $H_{U}$ and non-zero off diagonal element contributions from  $H_t$. In general, the elements of the Hamiltonian are given by $H_{nm} = \left\langle \alpha_n\right| H  \left|\alpha_m\right\rangle$. After determining the matrix elements $H_{nm}$ for a given gate voltage configuration, the  Hamiltonian is diagonalized to find the charge state associated with the ground state energy. This allows us to identify the charge state of a linear QD array as a function of the voltages applied to the gate electrodes.  

\section{Results and Discussion}

The following subsections 1) compare analytic and simulated results for onsite and inter-Coulomb energy calculated for a theoretical double QD electrostatic potential, and 2) present a charge stability diagram\index{Charge Stability Diagram} (CSD) calculated from realistic double QD electrostatic potentials. An electrostatic potential of two QDs separated by $d$ is approximated by a quartic function with a confinement strength parameter $\omega$:
\begin{equation}\label{eq:quartic_func}
    U(x,y) = \frac{m^* \omega^2}{2}\left[ \frac{(x^2 -\frac{d^2}{4})^2}{d^2} +y^2\right]
\end{equation}
and 2D Gaussian functions are used to approximate the single-particle ground state wave functions of each QD as described in \cite{Yang2011}. For the CSD, the electrostatic potential as a function of plunger gate voltages, $\boldsymbol{V} = [V_1, V_2]$, and tunnel gate voltages, $\boldsymbol{W} = [W_1]$, is calculated via finite-element Poisson solver \texttt{nextnano} \cite{Birner2007}, and then used as input to calculate the Hubbard model parameters and determine the charge state for a given voltage configuration. 

\subsection{Numerical Accuracy and Efficiency}
\label{sec:3.1}

\begin{figure}[b]
    \centering
    \begin{subfigure}[h!]{0.45\textwidth}
        \centering
        \includegraphics[width=\linewidth]{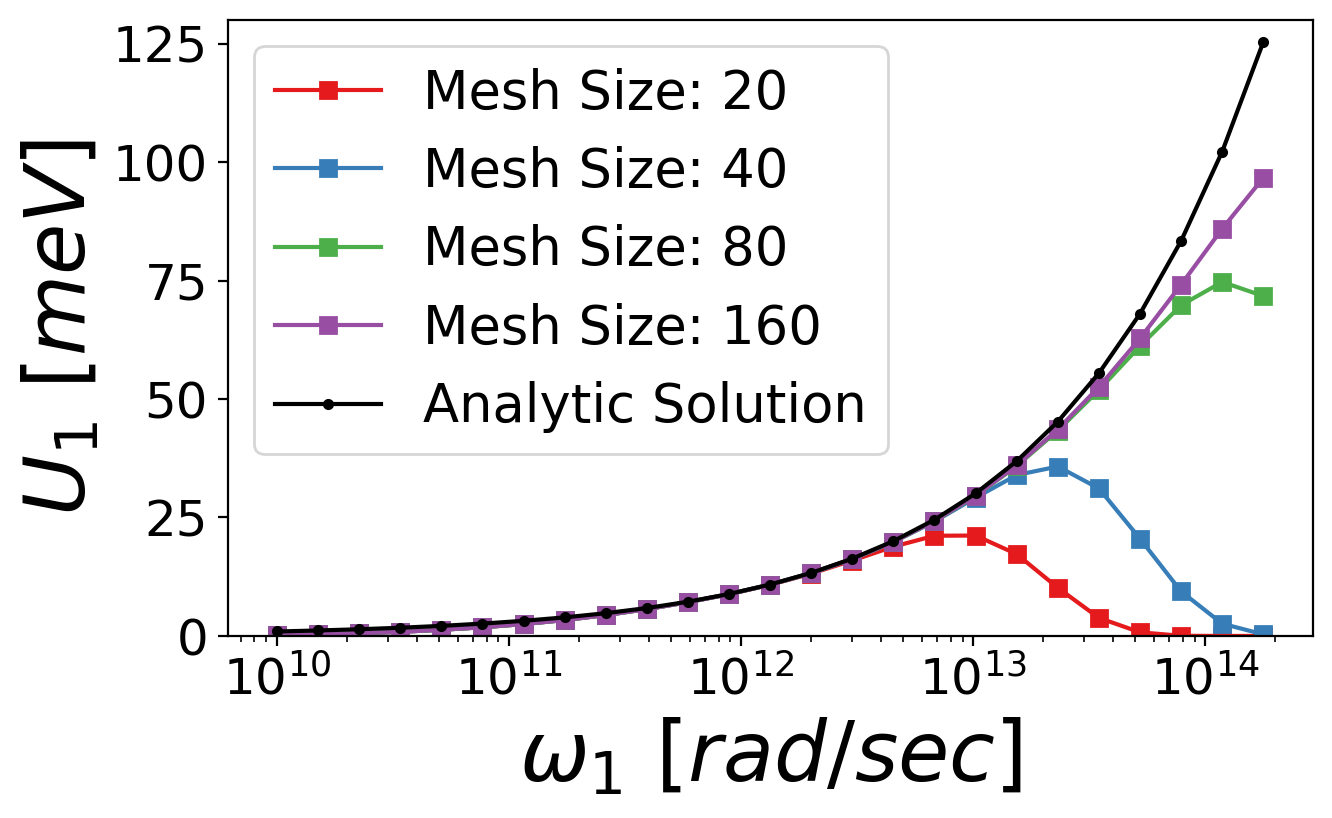} 
        \caption{} \label{fig:fig3a}
    \end{subfigure}
    \hspace{1em}%
    \begin{subfigure}[h!]{0.45\textwidth}
        \centering
        \includegraphics[width=\linewidth]{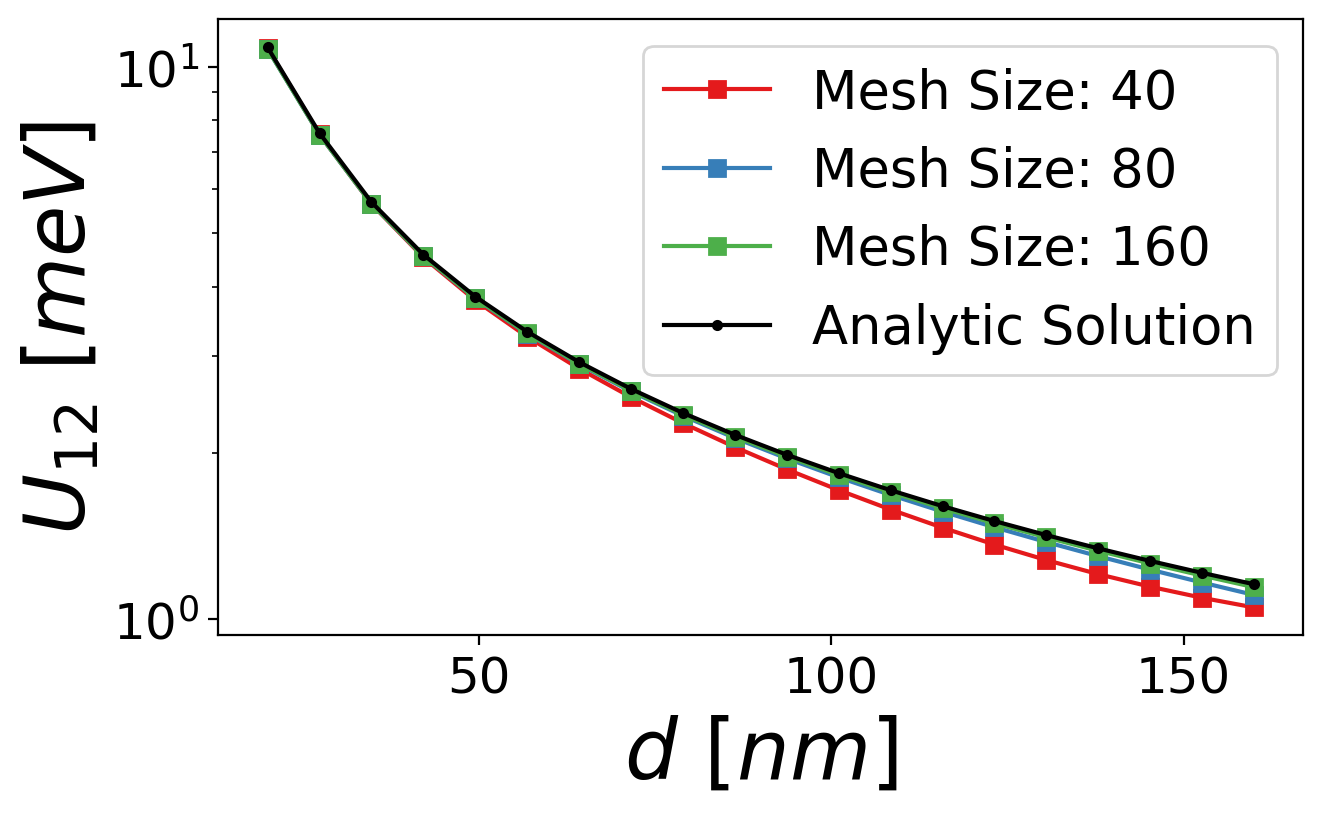} 
        \caption{} \label{fig:fig3b}
    \end{subfigure}
    \hspace{1em}%
    \begin{subfigure}[h!]{0.45\textwidth}
        \centering
        \includegraphics[width=\linewidth]{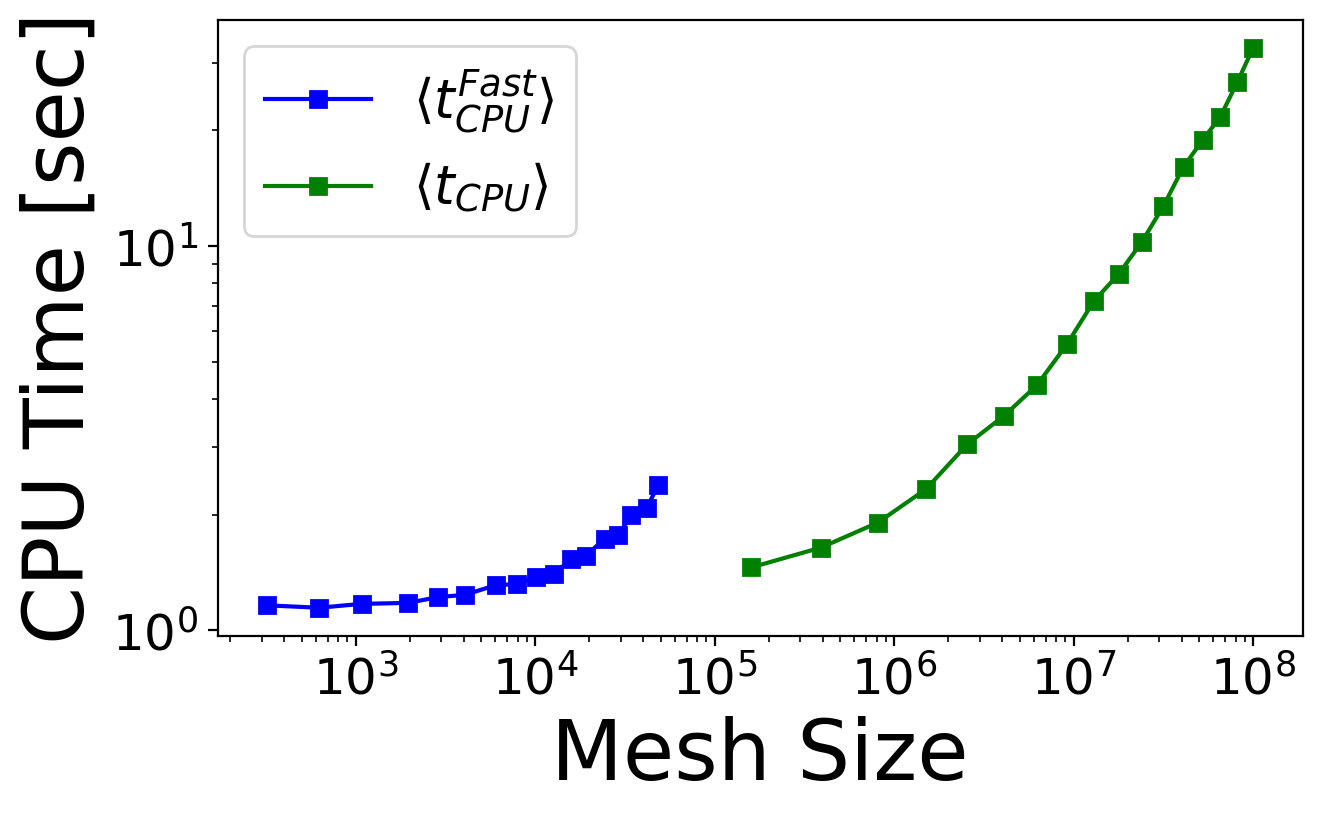} 
        \caption{} \label{fig:fig3c}
    \end{subfigure}
    \caption{Comparison of numerical integration results to analytic results based on a quartic electrostatic potential.  
        (\subref{fig:fig3a}) Onsite Coulomb energy $U_1$ versus QD confinement energy, $\omega_1$. 
        (\subref{fig:fig3b}) Interdot Coulomb energy $U_{12}$ versus dot separation, $d$. 
        (\subref{fig:fig3c}) Comparison of two integration methods, showing the average CPU time versus number of mesh grid points evaluated to obtain equivalent $U_1$ results. The data points labelled `fast' refer to an integration algorithm that evaluates mesh grid points only where the electron probability density is non-negligible.} \label{fig:fig3}
\end{figure}

Figure \ref{fig:fig3a} plots the onsite energy as a function of confinement strength (in angular frequency units), $\omega_1$, and compares the analytic solution to numerically obtained results with a varying number of mesh grid points. Note that the confinement strength $\omega_1$ is directly related to the QD radius, $r_0 = \sqrt{\hbar/(m^*\omega_1)}$. For weakly confined electrons (large QD), accurate results can be obtained using fewer mesh grid points, however, as the QD size decreases the number of mesh grid points must increase. 
Electrostatically-defined QDs typically have radii on the order of tens of nanometers, giving $\omega \approx 10^{11}$-$10^{12}$ rad/sec, which reduces the required number of mesh grid points and therefore the computation time compared to smaller QDs. Similarly, figure \ref{fig:fig3b} shows the inter-Coulomb energy as a function of QD separation parameter, $d$, for  two QDs with confinement $\omega_1 = \omega_2 = 5 \cdot 10^{12}$ rad/sec. A greater number of mesh grid points is required for numerical accuracy as interdot separation $d$ is increased, which is a feature of the mesh grid formalism used in the numerical polar integration.  

The numerical integration in (\ref{eq:Uij_polar}) involves a quadruple integral, with each variable being integrated over $n$ points within the coordinate boundaries defining the double QD electrostatic potential. Therefore, the number of mesh grid points to be evaluated during the numerical integration grows as $4^n$. Figure \ref{fig:fig3c} shows how the CPU time scales with the number of mesh grid points for the brute force numerical integration as well as an improved integration algorithm. The improved algorithm only evaluates mesh grid points in regions where the electron is likely to be located, which significantly reduces the overall mesh grid size and therefore the required CPU time.

\subsection{Charge Stability Diagram}
\label{sec:3.2}
The effective parameter calculations in section \ref{sec:3.1} were performed for a quartic electrostatic potential that was a function of two control parameters, i.e., the confinement strength $\omega$ and dot separation $d$. In realistic  devices, besides the chosen 3D geometry, the control parameters are the plunger, $\boldsymbol{V}$, and tunnel, $\boldsymbol{W}$, gate voltage vectors that govern the potential landscape.
\begin{figure}[htb!]
    \centering
    \begin{subfigure}[h!]{0.45\textwidth}
        \centering
        \includegraphics[width=\linewidth]{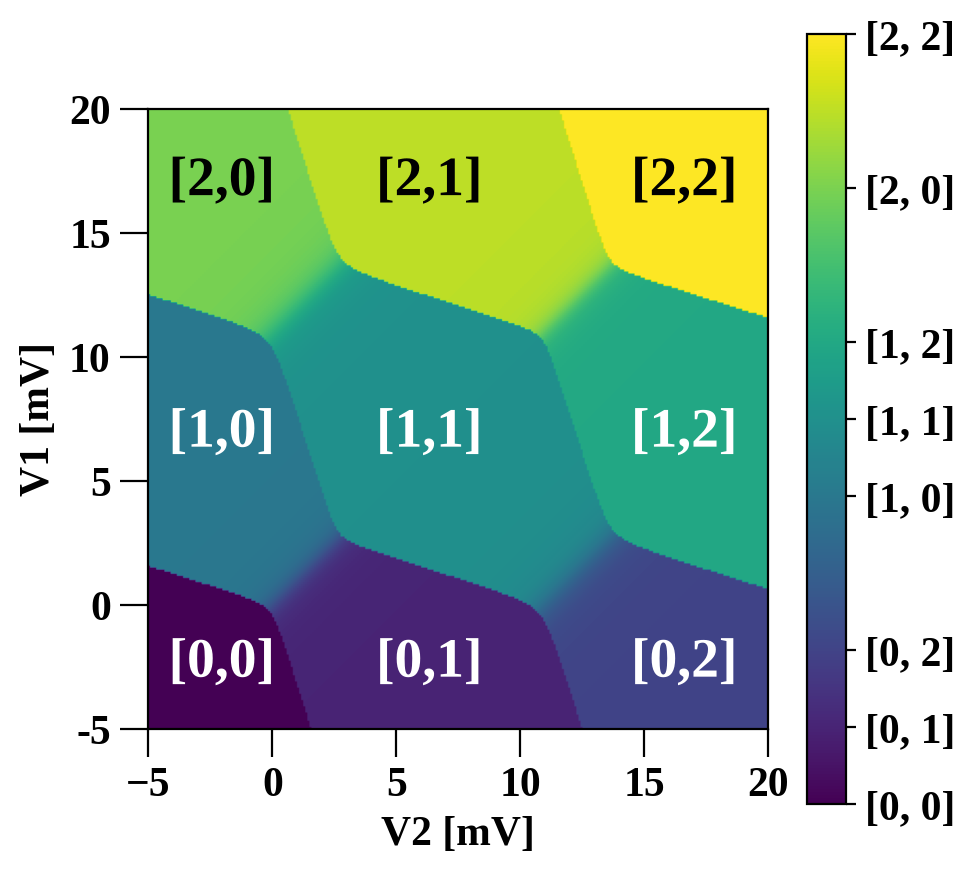} 
        \caption{} \label{fig:fig4a}
    \end{subfigure}
    \hspace{1em}%
    \begin{subfigure}[h!]{0.45\textwidth}
        \centering
        \includegraphics[width=\linewidth]{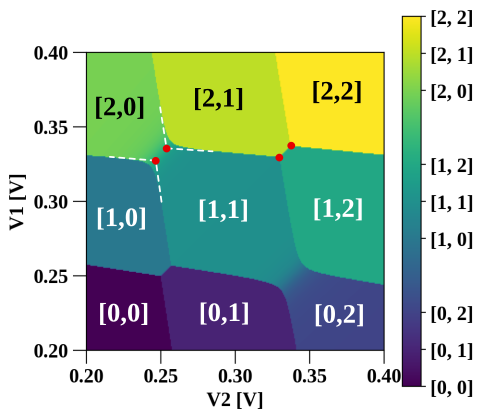} 
        \caption{} \label{fig:fig4b}
    \end{subfigure}
    \caption{Charge stability diagrams (CSD), where $[n,m]$ labels the electron occupation numbers of the two dots. (\subref{fig:fig4a}) Theoretical CSD based on a capacitive model in which $\mu_i(V_1, V_2)$ accounts for cross talk between QDs, while $t_{12}$, $U_{1}$, $U_{2}$, and $U_{12}$ are held constant. (\subref{fig:fig4b}) CSD resulting from the numerical Hubbard model approach. The gate voltage-dependent Hubbard parameters are calculated based on realistic QD potentials from a finite element Poisson solver. Two pairs of triple points (in red) are indicated; the left pair are in a regime of stronger tunnel coupling than the pair on the right. 
    } \label{fig:fig4}
\end{figure}
We determine the CSD for a realistic double QD device shown in \ref{fig: fig1c} by first constructing a charge basis consisting of 16 charge states with total charge ranging from 0 to 4 electrons, with the Hubbard model restriction that no more than 2 electrons can occupy each QD. The Hubbard Hamiltonian is calculated and diagonalized to find the charge state with lowest energy. This calculation is repeated for all simulated voltage configurations. 

In figure \ref{fig:fig4a}, a theoretical CSD is constructed assuming a capacitive model \cite{DasSarma2011} and quartic potential. In the model,  ${\mu}_{i}(V_1, V_2)$ are functions of the plunger gate voltages for a double QD device, and $t_{12}$, $U_{1}$, and $U_{12}$ are assumed to be constant and voltage-independent. Figure \ref{fig:fig4b} shows the CSD resulting from the numerical Hubbard model approach outlined in the present work applied to realistic double QD potentials, such as the one shown in figure \ref{fig: fig1c}. Although the voltage ranges of the plunger gates are different, the general features shown in figures \ref{fig:fig4a} and \ref{fig:fig4b} are comparable. Cross-capacitance effects are evident as the sloped charge transition lines. There are two primary advantages in using our Hubbard model approach with realistic QD potentials to determine the CSD in figure \ref{fig:fig4b}: 
\begin{enumerate}
    \item Subtle voltage dependencies in the onsite and inter-Coulomb energy are captured, which can be seen in figure \ref{fig:fig4b} by the varying charge transition line slopes and triple point separations.
    \item Strong voltage dependencies on the interdot tunnel coupling are captured, evident as variable smoothing of the charge transitions about the triple points for different regions of the CSD. 
\end{enumerate}

\section{Conclusion}
\label{sec:4}

In this work, an algorithmic approach has been presented for calculating the effective parameters of a Hubbard Hamiltonian that models electrons in realistic linear QD arrays. This was used to identify regions of charge stability as a function of control voltages  which captures features missed by other simplified models. Although the results presented are for a QD array of length two, the method generalizes to arbitrary-sized arrays. Further development is underway to perform larger QD array calculations for application to a node network surface code. These methods are a valuable tool within the larger toolbox of simulation methods applicable to the development of scalable spin qubit quantum processors. 

\bibliography{ammcs_proceedings}

\begin{thebibliography}{10}
\providecommand{\url}[1]{{#1}}
\providecommand{\urlprefix}{URL }
\expandafter\ifx\csname urlstyle\endcsname\relax
  \providecommand{\doi}[1]{DOI~\discretionary{}{}{}#1}\else
  \providecommand{\doi}{DOI~\discretionary{}{}{}\begingroup
  \urlstyle{rm}\Url}\fi

\bibitem{Birner2007}
Birner, S., Zibold, T., Andlauer, T., Kubis, T., Sabathil, M., Trellakis, A.,
  Vogl, P.: nextnano: General purpose 3-d simulations.
\newblock {IEEE} Transactions on Electron Devices \textbf{54}(9), 2137--2142
  (2007).
\newblock \doi{10.1109/ted.2007.902871}

\bibitem{Buonacorsi2019}
Buonacorsi, B., Cai, Z., Ramirez, E.B., Willick, K.S., Walker, S.M., Li, J.,
  Shaw, B.D., Xu, X., Benjamin, S.C., Baugh, J.: Network architecture for a
  topological quantum computer in silicon.
\newblock Quantum Science and Technology \textbf{4}(2), 025,003 (2019).
\newblock \doi{10.1088/2058-9565/aaf3c4}

\bibitem{DasSarma2011}
Das~Sarma, S., Wang, X., Yang, S.: Hubbard model description of silicon spin
  qubits: Charge stability diagram and tunnel coupling in si double quantum
  dots.
\newblock Physical Review B \textbf{83}(23) (2011).
\newblock \doi{10.1103/physrevb.83.235314}

\bibitem{Heinz2021}
Heinz, I., Burkard, G.: Crosstalk analysis for single-qubit and two-qubit gates
  in spin qubit arrays.
\newblock Physical Review B \textbf{104}(4), 045,420 (2021).
\newblock \doi{10.1103/physrevb.104.045420}

\bibitem{Hensgens2017}
Hensgens, T., Fujita, T., Janssen, L., Li, X., Van~Diepen, C., Reichl, C.,
  Wegscheider, W., Das~Sarma, S., Vandersypen, L.: Quantum simulation of a
  fermi–hubbard model using a semiconductor quantum dot array.
\newblock Nature \textbf{548}(7665), 70--73 (2017).
\newblock \doi{10.1038/nature23022}

\bibitem{Jafari2008}
Jafari, S.A.: Introduction to hubbard model and exact diagonalization.
\newblock {arXiv}  (2008).
\newblock \doi{10.48550/ARXIV.0807.4878}.
\newblock \urlprefix\url{https://arxiv.org/abs/0807.4878}

\bibitem{Mills2022}
Mills, A.R., Guinn, C.R., Gullans, M.J., Sigillito, A.J., Feldman, M.M.,
  Nielsen, E., Petta, J.R.: Two-qubit silicon quantum processor with operation
  fidelity exceeding 99{\%}.
\newblock Science Advances \textbf{8}(14) (2022).
\newblock \doi{10.1126/sciadv.abn5130}

\bibitem{Petit2020}
Petit, L., Eenink, H.G.J., Russ, M., Lawrie, W.I.L., Hendrickx, N.W., Philips,
  S.G.J., Clarke, J.S., Vandersypen, L.M.K., Veldhorst, M.: Universal quantum
  logic in hot silicon qubits.
\newblock Nature \textbf{580}(7803), 355--359 (2020).
\newblock \doi{10.1038/s41586-020-2170-7}

\bibitem{Saraiva2021}
Saraiva, A., Lim, W.H., Yang, C.H., Escott, C.C., Laucht, A., Dzurak, A.S.:
  Materials for silicon quantum dots and their impact on electron spin qubits.
\newblock Advanced Functional Materials \textbf{32}(3) (2021).
\newblock \doi{10.1002/adfm.202105488}

\bibitem{Secchi2023}
Secchi, A., Troiani, F.: Theory of multidimensional quantum capacitance and its
  application to spin and charge discrimination in quantum dot arrays.
\newblock Physical Review B \textbf{107}(15) (2023).
\newblock \doi{10.1103/physrevb.107.155411}

\bibitem{Undseth2023}
Undseth, B., Xue, X., Mehmandoost, M., Rimbach-Russ, M., Eendebak, P.T.,
  Samkharadze, N., Sammak, A., Dobrovitski, V.V., Scappucci, G., Vandersypen,
  L.M.: Nonlinear response and crosstalk of electrically driven silicon spin
  qubits.
\newblock Physical Review Applied \textbf{19}(4), 044,078 (2023).
\newblock \doi{10.1103/physrevapplied.19.044078}

\bibitem{Veldhorst2017}
Veldhorst, M., Eenink, H.G.J., Yang, C.H., Dzurak, A.S.: Silicon {CMOS}
  architecture for a spin-based quantum computer.
\newblock Nature Communications \textbf{8}(1) (2017).
\newblock \doi{10.1038/s41467-017-01905-6}

\bibitem{Veldhorst2014}
Veldhorst, M., Hwang, J.C.C., Yang, C.H., Leenstra, A.W., de~Ronde, B.,
  Dehollain, J.P., Muhonen, J.T., Hudson, F.E., Itoh, K.M., Morello, A.,
  Dzurak, A.S.: An addressable quantum dot qubit with fault-tolerant
  control-fidelity.
\newblock Nature Nanotechnology \textbf{9}(12), 981--985 (2014).
\newblock \doi{10.1038/nnano.2014.216}

\bibitem{Yang2011}
Yang, S., Wang, X., Das~Sarma, S.: Generic hubbard model description of
  semiconductor quantum-dot spin qubits.
\newblock Physical Review B \textbf{83}(16) (2011).
\newblock \doi{10.1103/physrevb.83.161301}

\bibitem{Zhong2020}
Zhong, H.S., Wang, H., Deng, Y.H., Chen, M.C., Peng, L.C., Luo, Y.H., Qin, J.,
  Wu, D., Ding, X., Hu, Y., Hu, P., Yang, X.Y., Zhang, W.J., Li, H., Li, Y.,
  Jiang, X., Gan, L., Yang, G., You, L., Wang, Z., Li, L., Liu, N.L., Lu, C.Y.,
  Pan, J.W.: Quantum computational advantage using photons.
\newblock Science \textbf{370}(6523), 1460--1463 (2020).
\newblock \doi{10.1126/science.abe8770}

\end{thebibliography}

\end{document}